\documentclass[runningheads,a4paper]{llncs}

\usepackage{amssymb}
\usepackage{graphicx}

\usepackage{amsmath}
\usepackage{stfloats}
\usepackage{amsfonts}

\usepackage{url}
\urldef{\mailsa}\path|{alfred.hofmann, ursula.barth, ingrid.haas, frank.holzwarth,|
\urldef{\mailsb}\path|anna.kramer, leonie.kunz, christine.reiss, nicole.sator,|
\urldef{\mailsc}\path|erika.siebert-cole, peter.strasser, lncs}@springer.com|
\newcommand{\keywords}[1]{\par\addvspace\baselineskip
\noindent\keywordname\enspace\ignorespaces#1}

\begin{document}

\mainmatter  

\title{The Cut and Dominating Set Problem in A Steganographer Network}

\titlerunning{The Cut and Dominating Set Problem in A Steganographer Network}

%
%
\author{
Hanzhou Wu$^{1\dagger}$, Wei Wang$^{1\ddagger}$, Jing Dong$^{1\ddagger}$, Hongxia Wang$^2$ and Lizhi Xiong$^3$}
\authorrunning{H. Wu et al.}

\institute{$^1$Institute of Automation, Chinese Academy of Sciences, Beijing 100190, China\\
$^2$School of Inf. Sci. \& Technol., Southwest Jiaotong U., Chengdu 611756, China\\
$^3$School of Comp. \& Softw., Nanjing U. of Inf. Sci. \& Technol., Nanjing 210044, China\\
\email{$^\dagger$h.wu.phd@ieee.org, $^\ddagger$\{wwang,jdong\}@nlpr.ia.ac.cn}
}
%
%

\maketitle

\begin{abstract}
A steganographer network corresponds to a graphic structure that the involved vertices (or called nodes) denote social entities such as the data encoders and data decoders, and the associated edges represent any real communicable channels or other social links that could be utilized for steganography. Unlike traditional steganographic algorithms, a steganographer network models steganographic communication by an abstract way such that the concerned underlying characteristics of steganography are quantized as analyzable parameters in the network. In this paper, we will analyze two problems in a steganographer network. The first problem is a passive attack to a steganographer network where a network monitor has collected a list of suspicious vertices corresponding to the data encoders or decoders. The network monitor expects to break (disconnect) the steganographic communication down between the suspicious vertices while keeping the cost as low as possible. The second one relates to determining a set of vertices corresponding to the data encoders (senders) such that all vertices can share a message by neighbors. We point that, the two problems are equivalent to the minimum cut problem and the minimum-weight dominating set problem.
\end{abstract}

\section{Introduction}
Steganography \cite{cox:book} has been extensively studied in past twenty years. As a means to secret communication, different from cryptography, steganography refers to the art of hiding a message into an innocent digital object also called \emph{cover} by slightly modifying the noise-like components of the host. The resulting object also called \emph{stego} will not introduce any noticeable artifacts and will be sent to the receiver. As steganography even conceals the presence of communication, it serves an important role in information security nowadays.

A number of advanced steganographic algorithms and novel perspectives have been reported in the literature such as \cite{holub:suniward}, \cite{guo:uedr}, \cite{li:tifs}, \cite{tang:gan}, \cite{zhou:reassign}, \cite{wu:isbast}, \cite{wu:mtap}, \cite{jessica:stcs}, \cite{zhao:spe}. A most important requirement \cite{wu:mtap} for any reliable steganographic system is that it should be impossible (or say very hard) for an eavesdropper to distinguish between ordinary objects and objects containing hidden information. The mainstream design concept of steganography is to minimize the data embedding impact on the cover for a required payload, which involves at least two aspects for achieving superior steganographic performance. The first is to find the most suitable cover elements for data embedding. For example, complex texture areas within an image would be quite desirable for steganography. The second is to maximize data embedding efficiency. For the former, nowadays, we often focus on designing a suited function for the cover elements that can expose the data embedding impact. For the latter, cover codes are the key technologies.

Steganography can be also utilized for malicious purposes, which gives the life of another important research topic called steganalysis. A core work in steganalysis is to detect whether a given digit object was embedded with a message or not. The conventional steganalysis algorithms regard this problem as a binary classification task, which involves feature engineering and classifiers. Due to the advancement of steganography that tends to alter the cover regions that are quite hard to detect, conventional featured-based steganalysis algorithms relying on sophisticated manual feature design has been pushed to the limit and become very hard to improve. To overcome this difficulty, in-deep studies can be performed on moving the success achieved by deep convolutional neural networks (deep CNNs) in computer vision to steganalysis \cite{xu:sdcnn}, \cite{xu:ensemble}, \cite{xu:cnnJUNIWARD}, \cite{ye:cnn}. It is pointed that, estimating the payload size within a stego, extracting the embedded information and identifying steganographic senders/receivers are all the important yet quite challenging research topics in steganalysis.

As mentioned in \cite{hzwu:cihw}, with the rapidly development of social media services such as Facebook and Twitter, it would be quite desirable to study the social behaviors, protocols and any other scenarios of steganography. In this sense, we may not care about the details of the used steganographic algorithms, but rather quantize the concerned characteristics of steganography as analyzable parameters for problem optimization. An effective way to analyze steganographic activities is to model it in a social network, which consists of a set of vertices and a set of edges. The vertices represent social entities such as the data encoders and data decoders, and the associated edges represent any real communicable channels or other social links used for steganography. In \cite{hzwu:cihw}, the authors model steganography in an additive-risk social network and present a communication strategy that aims to minimize the steganographic risk by choosing a set of edges with the minimum sum of weights.

In this paper, we present a passive attack to steganography in a steganographer network. We assume that the attacker serves as a network monitor, and has collected a list of suspicious vertices that are corresponding to the data encoders and data decoders in advance. Here, the ``suspicious vertices'' mean that they are with a high probability representing the steganographic actors though they may be misjudged in practice. One may use the conventional steganalysis algorithms or other anomaly detection algorithms to construct the suspicious vertex-set. We study such a passive attack that, the network monitor hopes to remove a set of edges so that the suspicious vertices cannot communicate via steganography while the removal cost could be minimized. In addition to the above-mentioned attack, we also analyze a new communication scenario in the steganographer network that all the vertices will serve as either a data encoder or a data decoder. The optimization task is to select a set of vertices (encoders) out such that all vertices can share a message by edges and the steganographic communication activities are performed between \emph{adjacent} vertices in the network. Similarly, we expect to keep the vertex-selection risk as low as possible.

The rest of this paper are organized as follows. The related work is presented in Section 2. In Section 3, we introduce the proposed passive attack. We analyze a new communication scenario in a steganographer network in Section 4. Finally, we conclude this paper in Section 5.
\begin{figure}[!t]
\centering
\includegraphics[width=2.4in]{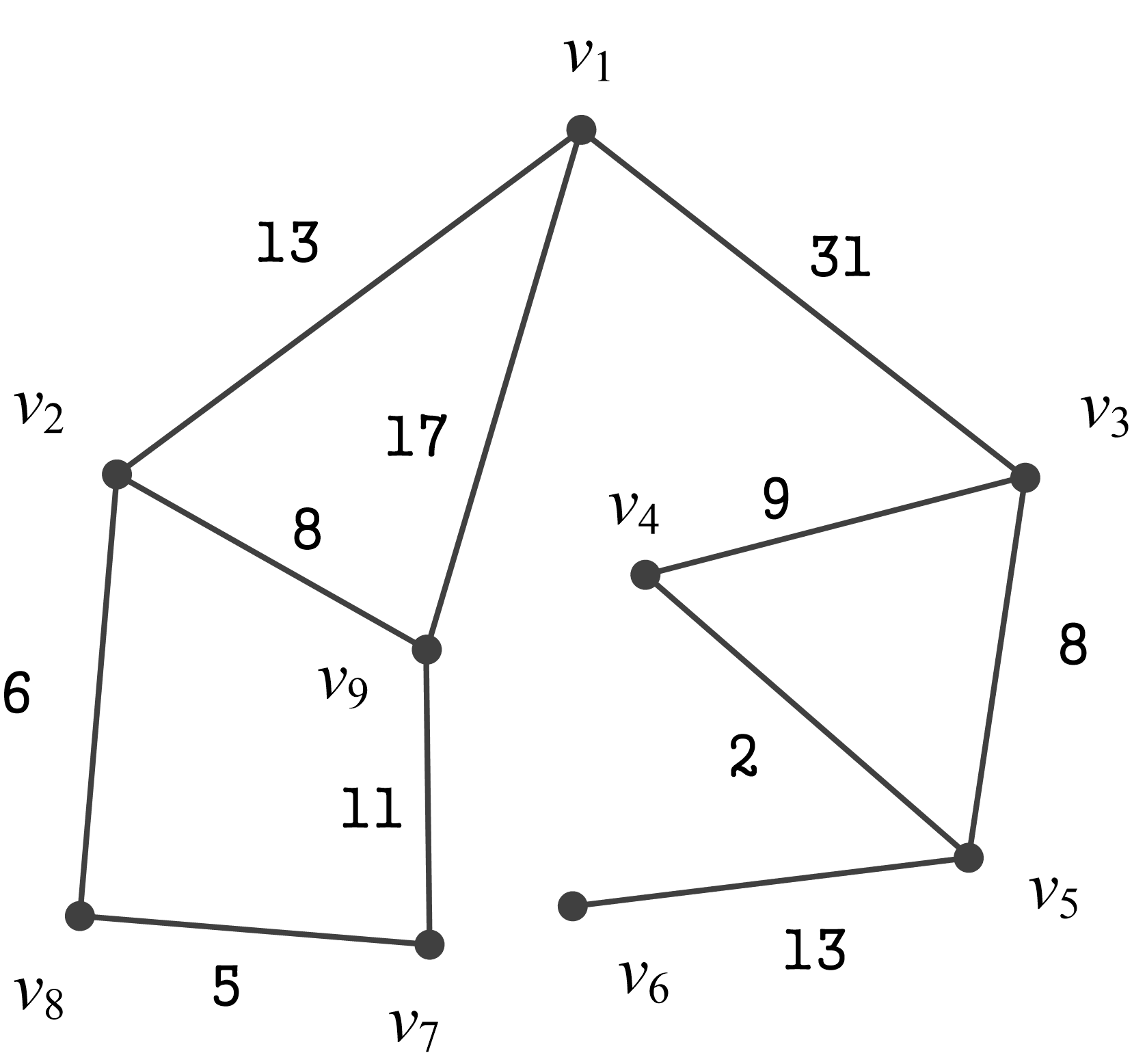}
\caption{An example for steganographer network where the weights represent the corresponding removel costs.}
\end{figure}

\section{Related Work}
A steganographer network corresponds to a graph $G(V, E)$, where $V = \{v_1, v_2,$ $..., v_n\}$ represents the set of vertices and $E = \{e_1, e_2, ..., e_m\}$ represents the set of edges. Every edge $e_k\in E$ involves with a pair of vertices, i.e., $e_k = (v_i,v_j)$. We say $G$ is undirected meaning that all $e\in E$ have no orientation. Namely, $(v_i,v_j)$ is equivalent to $(v_j,v_i)$. A path (if any) between $v_i$ and $v_j$ corresponds to such a vertex sequence $(v_{q_1}, v_{q_2}, ..., v_{q_t})$ that $v_{q_1} = v_i$, $v_{q_t} = v_j$ and $(v_{q_{k-1}}, v_{q_k})\in E$ for all $2\leq k\leq t$. $G$ is connected if and only if for any two vertices in $G$, there exists at least one path between them.

In $G$, a path between $v_i$ and $v_j$ implies that, $v_i$ can share a message with $v_j$ along the path. In \cite{hzwu:cihw}, the authors present a communication strategy that aims to minimize the overall steganographic risk by determining a subtree supporting the required vertices. Mathematically, they use $S = \{s_1, s_2,$ $..., s_{n_1}\}$ $\subset V$ and $T = \{t_1, t_2,$ $..., t_{n_2}\}$ $\subset V$ to denote the encoder set and decoder set. For each $s_i \in S$, its individual decoder set is denoted by $T_i\subset T$, meaning that, $s_i$ hopes to send a message to each of $T_i$. Here, $\cup_{i=1}^{n_1}T_i = T$. The optimization task is to find such a subset of $E$ that it enables all $s_i\in S$ to send a message to all decoders in $T_i$, and the overall risk can be minimized. Namely,
\begin{equation}
E_{\textrm{opt}}(S, T) = \underset{E_{\textrm{usable}}\subset E}{\textrm{arg min}}~~~R(S, T, E_\textrm{usable}),
\end{equation}
where $R(S, T, E_\textrm{usable})$ denotes the risk over $E_\textrm{usable}\subset E$.

By assuming an additive-risk network, the optimization task is defined as:
\begin{equation}
E_{\textrm{opt}}(S, T) = \underset{E_{\textrm{usable}}\subset E}{\textrm{arg min}}~\sum_{e_i\in E_{\textrm{usable}}}w_i,
\end{equation}
where $w_i>0$ is a real number representing the risk of $e_i$.
\begin{figure}[!t]
\centering
\includegraphics[width=4.2in]{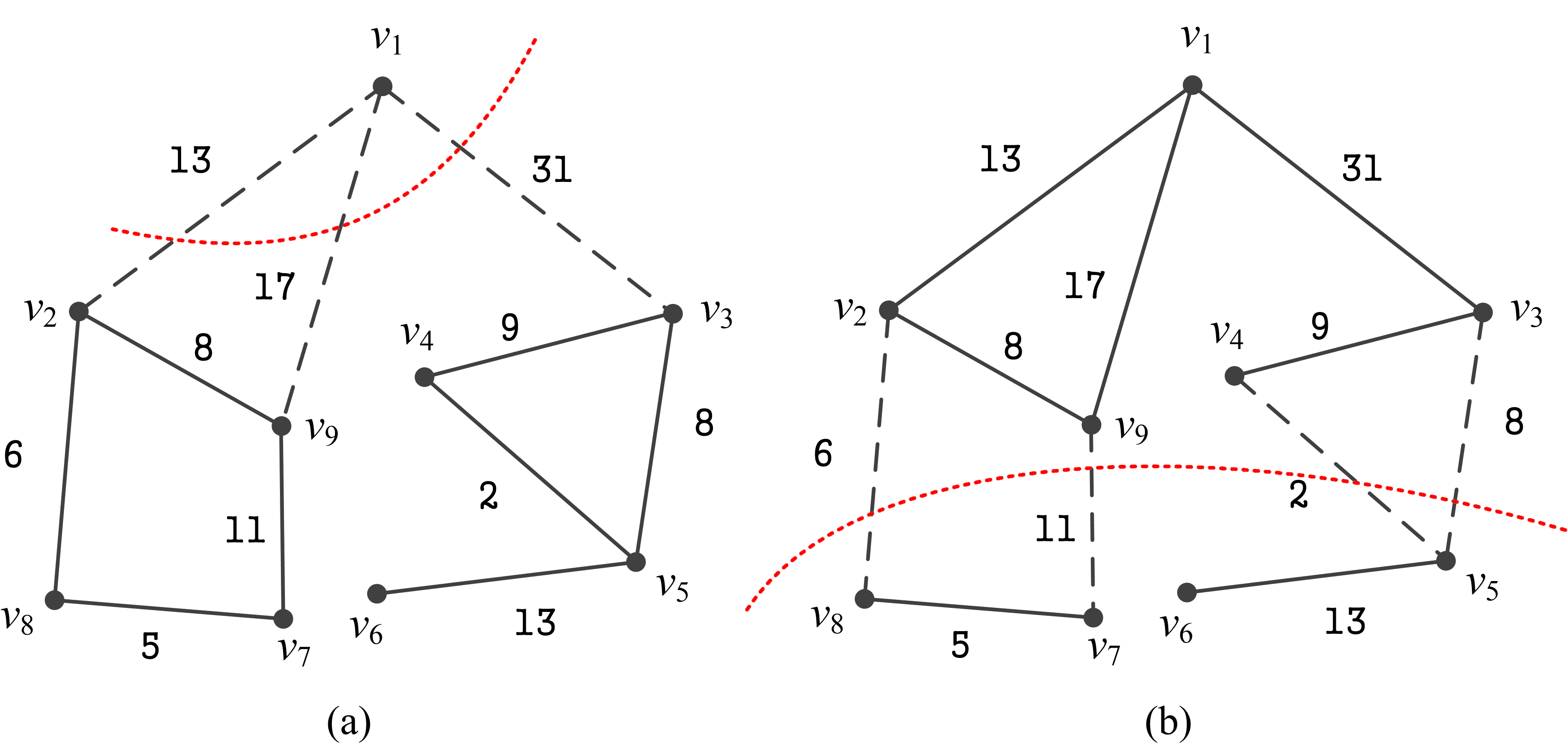}
\caption{Two ways to remove a set of edges making $S$ and $T$ unconnected where $S=\{v_1\}$ and $T=\{v_5,v_7,v_8\}$.}
\end{figure}

The authors in \cite{hzwu:cihw} have pointed that Eq. (2) is equivalent to solving the Steiner Tree Problem (STP) problem \cite{chlebik:stp}, which is NP-hard. The STP is seen as a generalization of two other combinatorial optimization problems, i.e., the shortest path problem and the minimum spanning tree (MST) problem. If the STP deals with only two terminals, it reduces to finding a shortest path. If all vertices are terminals, it is equivalent to the MST problem. Therefore, one may employ the existing approximation algorithms designed for the STP to find the suitable strategy for steganographic communication.

The authors also analyze the steganographic communication from a probabilistic graph perspective. They have proved that, a multiplicative probabilistic graph is equivalent to an additive weighted graph, meaning that, by translating the probabilistic network into a weighted network, one could find the suited communication strategy in a similar way.

\section{Passive Attack to Steganographer Network}
From a monitor point, even if a steganographer does not use the optimized communication strategy, he can still communicate a message with the desired receiver since there may exist multiple communicable paths in the network. This requires the monitor to remove a set of edges corresponding to the channels such that the steganographer cannot communicate with the receiver. However, edge removal may result in a high cost or risk. We study this problem in this section. Mathematically, we expect to remove a set of edges $E_\text{opt}$ such that $S$ cannot communicate with $T$ via steganography, while the removal operation could result in the lowest risk (cost), i.e.,
\begin{equation}
E_{\textrm{opt}}(S, T) = \underset{E_{\textrm{rem}}\subset E}{\textrm{arg min}}~~~R(S, T, E_\textrm{rem}).
\end{equation}

We assign a weight $w_i > 0$ to $e_i\in E$. $w_i$ represents the cost of removing $e_i$ from $G$. We assume an additive $G$, i.e.,
\begin{equation}
R(S, T, E_\textrm{rem}) = \sum_{e_i\in E_\text{rem}}~~~w_i,
\end{equation}
which makes the problem amenable to mathematical analysis and the additive assumption is reasonable since two edges far away to each other have ignorable interaction and interaction between edges near to each other can be scored by weights.
\begin{figure}[!t]
\centering
\includegraphics[width=3.2in]{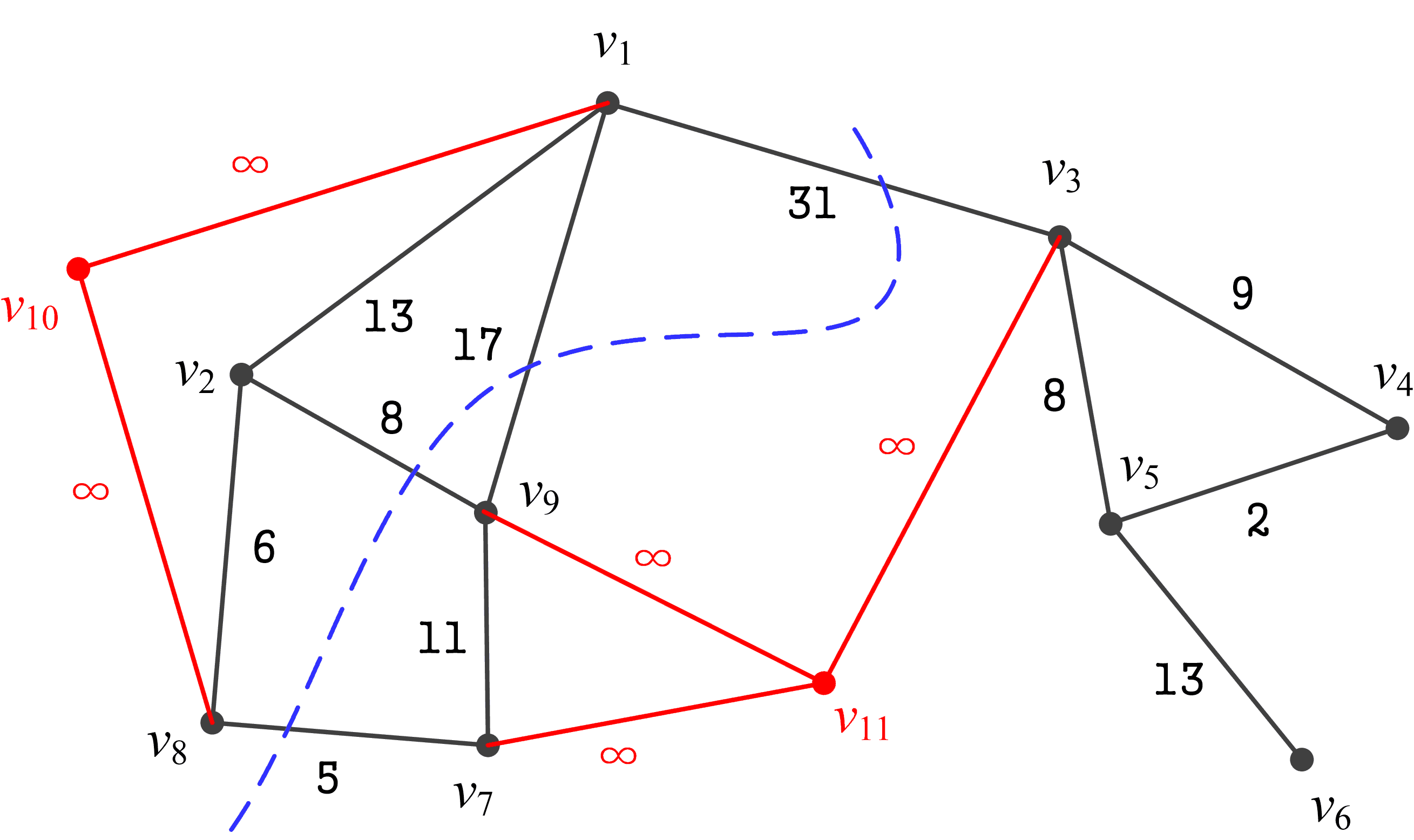}
\caption{Vertex insertion and edge insertion in a steganographer network.}
\end{figure}

In the real-world, for the network monitor, it is very likely that he cannot identify $T_i$ for $s_i$, meaning that, he may aim to disconnect the links between $S$ and $T$, rather than that between $s_i$ and $T_i$ for all $1\leq i\leq n_1$. This assumption has been utilized in this section. We define the simplest attack for which $|S|=1$ as \emph{single encoder attack}. We expect to find a subset of $E$ such that there has no path between $s_1$ and each vertex in $T$ by removing the selected edges. And, the sum of weights should be minimum. Fig. 1 shows a steganographer network. Assuming that, $S = \{v_1\}$ and $T = \{v_5, v_7, v_8\}$. Fig. 2 provides two ways to divide $S$ and $T$ into two different connected components. It is seen that, different sets of edges corresponds to different costs, e.g., the cost for Fig. 2 (a) is $13+17+31=61$, and that for Fig. 2 (b) is $27$. We should find a method to separate $S$ and $T$ with the lowest cost. Notice that, we assume that $S\cap T=\emptyset$.

In graph theory, a \emph{cut} is defined as a partition of the vertices of a graph into two disjoint subsets. Specifically, a cut $C=(A,B)$ divides $V$ into two subsets $A$ and $B$. The \emph{cut-set} of $C$ corresponds to an edge-set $\{(u,v)\in E|u\in A, v\in B\}$. Notice that, $A\cap B=\emptyset$ and $A\cup B = V$. If $a\in V$ and $b\in V$ are specified vertices of $G$, we then say \emph{a-b cut} is a cut in which $a\in A$ and $b\in B$. In a weighted graph, the weight of a cut is defined as the sum of weights of edges belonging to the cut. A cut is therefore called as \emph{minimum cut} if its weight is not larger than that of any other cut. Obviously, the minimum cut is not unique.

A \emph{directed} graph is called as a flow network if each edge has a capacity and receives a flow. A flow in a directed graph must satisfy the restriction that the amount of flow into a vertex is equal to the amount of flow out of the vertex, unless it is the source vertex which has only outgoing flow, or sink vertex which has only incoming flow. Mathematically, the capacity of a directed edge $v_i\rightarrow v_j$ is a mapping $c: E\mapsto \mathbb{R}^{+}$, denoted by $c(v_i,v_j)$. It represents the maximum amount of flow that can pass through $v_i\rightarrow v_j$. It is noted that, $c(v_j,v_i)$ is different from $c(v_i,v_j)$. If $v_i\rightarrow v_j\notin E$, one may consider $c(v_i,v_j) = 0$. A flow is a mapping $f: E\mapsto \mathbb{R}^{+}$ subject to two constraints \cite{algo:2009}:
\begin{equation}
f(v_i,v_j) \leq c(v_i,v_j),\forall~v_i\rightarrow v_j \in E,
\end{equation}
and
\begin{equation}
\sum_{v_i\rightarrow v_j \in E}f(v_i,v_j) = \sum_{v_j\rightarrow v_i \in E}f(v_j,v_i), \forall~v_i \in V\setminus \{a, b\},
\end{equation}
where $a$ is the source vertex and $b$ is the sink vertex.

Without the loss of generality, an \emph{undirected} graph can be treated as ``directed'' since an \emph{undirected} edge $(v_i,v_j)\in E$ can be decomposed into two \emph{directed} edges $v_i\rightarrow  v_j$ and $v_j \rightarrow  v_i$. For simplicity, we sometimes use $v_i\rightarrow  v_j$ to represent a directed edge from $v_i$ to $v_j$ and $v_j \rightarrow  v_i$ from $v_j$ to $v_i$ though $(v_i,v_j)\in E$ is undirected in $G$ in this paper. It also means that, $v_i\rightarrow  v_j \in E$ and $v_j\rightarrow  v_i \in E$.

With Eqs. (5, 6), the value of a flow is defined by $|f| = \sum_{v\in V}f(a, v)$ or $|f| = \sum_{v\in V}f(v, b)$. It means the amount of flow from the source vertex to the sink vertex. The \emph{maximum flow problem} is therefore to maximize $|f|$, namely, to route as much flow as possible from $a$ to $b$. Notice that, if $a\rightarrow v\notin E$, one may set $f(a,v)=0\leq c(a,v) = 0$. A lot of practical algorithms can be applied for determining the maximum flow such as \cite{ff:1956}, \cite{ek:acm}, \cite{dinic:mf}. For a steganographer network, by regarding the weights as the flow capacity, we can determine the maximum flow between any two vertices. Therefore, an undirected steganographer network can be translated as a flow network. A necessary preprocessing is to decompose each edge into two directed edges, for which both the newly assigned weights should equal the original one.
\begin{figure}[!t]
\centering
\includegraphics[width=4.5in]{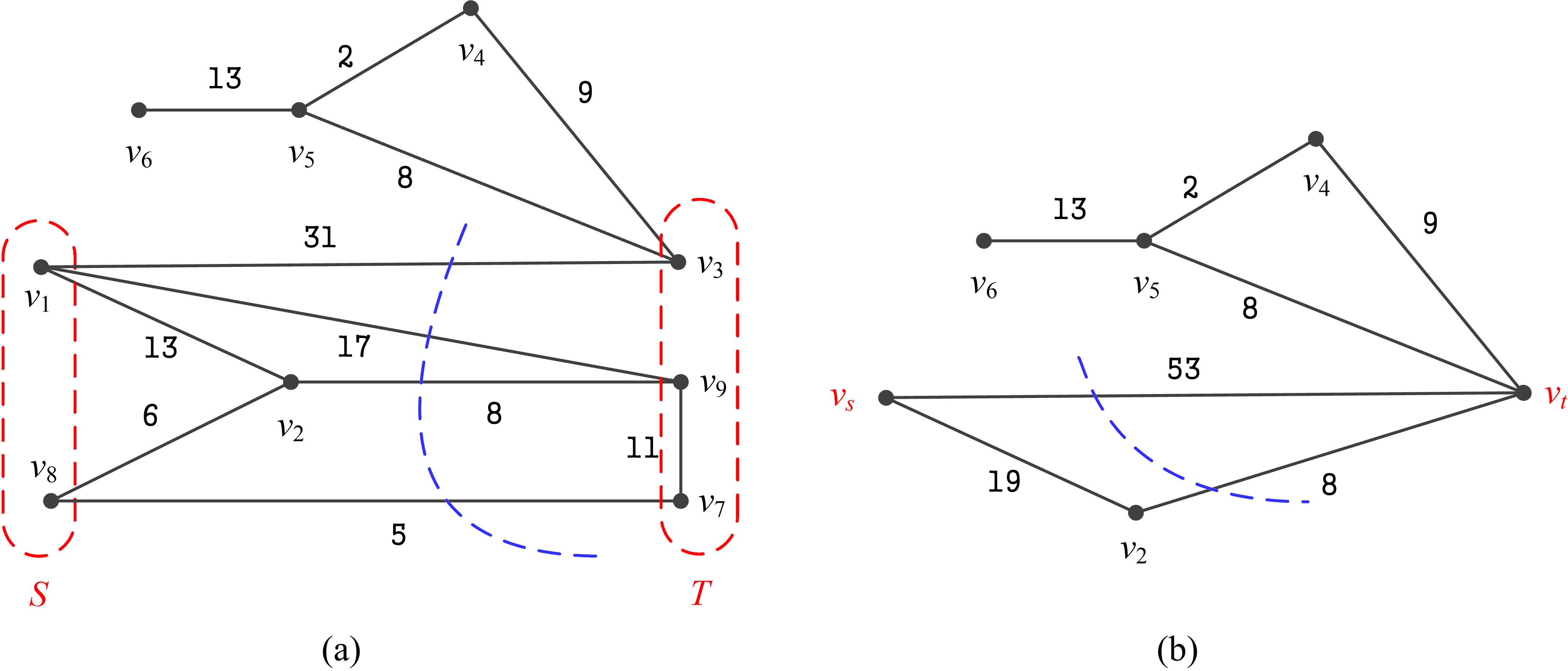}
\caption{Vertex contraction in a steganographer network: (a) vertex contraction and (b) a new graph.}
\end{figure}

The \emph{max-flow min-cut theorem} \cite{algo:2009} states that,
\begin{theorem}
In a flow network, the maximum amount of flow passing through the source vertex is equal to the weight of the minimum cut.
\end{theorem}

It indicates that, the smallest weight sum of the edges which if removed would disconnect the source vertex from the sink vertex is equal to the weight of the minimum cut, which is also equal to the maximum flow from the source to the sink. In order to find $E_{\textrm{opt}}(S, T)$ shown in Eq. (3), we need to determine the maximum flow in $G(V,E)$. Then, we construct the minimum cut. The edges in the minimum cut will constitute $E_{\textrm{opt}}(S, T)$. The maximum flow problem generally deals with only one source vertex and one sink vertex. However, we have $n_1\geq 1$ or $n_2\geq 1$, indicating that, we cannot directly use a maximum flow algorithm in the steganographer network since there may be multiple source vertices, i.e., $s_1, s_2, ..., s_{n_1}$, or multiple sink vertices, i.e., $t_1, t_2, ..., t_{n_2}$.

A way to address this issue is to insert a new \emph{super-source} vertex $v_s$ and a \emph{super-sink} vertex $v_t$ into $G(V,E)$, i.e.,
\begin{equation}
V = V\cup \{v_s, v_t\}.
\end{equation}

For each $s_i\in S$, we insert an edge $(v_s, s_i)$ into $E$. Then, for each $t_i\in T$, we insert $(t_i, v_t)$ into $E$. Therefore,
\begin{equation}
E = E\cup \{(v_s, s_i)|1\leq i\leq n_1\}\cup \{(t_i,v_t)|1\leq i\leq n_2\}.
\end{equation}

Thereafter, we assign an infinite large weight to each of the new edges to represent the cost of removing it from the new graph. We take Fig. 3 for explanation. In Fig. 3, we have $S=\{v_1, v_8\}$ and $T=\{v_3, v_7, v_9\}$. A super-source vertex $v_{10}$ (i.e., $v_s$) and a super-sink vertex $v_{11}$ (i.e., $v_t$) are inserted. Five new edges with the infinite large weight are inserted as well. In applications, one could assign a very large weight (that is larger than the sum of weights of all original edges) to the new edges, instead of the infinite large weight. Accordingly, a new graph can be built, as shown in Fig. (3). To find the maximum flow from $v_{10}$ to $v_{11}$, we should decompose all edges shown in Fig. (3) into two directed edges. By treating the weights of edges as the flow capacity, the maximum flow from $v_{10}$ to $v_{11}$ can be determined as 61. And, the corresponding minimum cut is $\{(v_1, v_3), (v_1, v_9), (v_2, v_9), (v_8, v_7)\}$. It means that, the minimum cost making $S$ and $T$ unconnected is 61, and $E_{\textrm{opt}}(S, T) = \{(v_1, v_3), (v_1, v_9), (v_2, v_9), (v_8, v_7)\}$.

Another way to determine $E_{\textrm{opt}}(S, T)$ is vertex contraction. The contraction of a set of vertices produces a graph in which all vertices in the set are replaced with a single vertex such that the single vertex is adjacent to the union of the vertices to which the vertices in the set were originally adjacent. In vertex contraction, it does not matter if two vertices in $S$ or $T$ are connected by an edge, which (if any) is simply removed during contraction.

In detail, we initialize $G'(V',E')$ as $V'=\emptyset$ and $E'=\emptyset$. For each $(v_i,v_j)\in E$, we skip it if $\{v_i,v_j\}\subset S$ or $\{v_i,v_j\}\subset T$. Otherwise, if $v_i\notin S\cup T$ and $v_j\notin S\cup T$, we update $V'$ as $V'=V'\cup \{v_i,v_j\}$ and $E'$ as $E'=E'\cup \{(v_i,v_j)\}$. The weights assigned to edges should be processed as well. If $v_i\in S$ and $v_j\in T$, we first update $V'$ as $V'=V'\cup \{v_s, v_t\}$ and then update $E'=E'\cup \{(v_s, v_t)\}$. The weight of $(v_i,v_j)$ will be added to $(v_s, v_t)$. If $v_i\in S$ and $v_j\notin T$, we first update $V'$ as $V'=V'\cup \{v_s, v_j\}$ and then update $E'=E'\cup \{(v_s, v_j)\}$. The weight of $(v_i,v_j)$ will be added to $(v_s, v_j)$. It is similar to process other cases. In this way, a new graph can be finally constructed. Thereafter, by computing the maximum flow from $v_s$ to $v_t$ in the new graph, the minimum cost can be obtained, and the minimum cut can be constructed as well. Fig. 4 shows an example, in which $\{v_1, v_8\}$ and $\{v_3, v_7, v_9\}$ are replaced with $v_s$ and $v_t$, respectively. It is seen that, the maximum flow in the new graph is identical to the maximum flow shown in Fig. 3, which has verified the correctness. Notice that, in Fig. 4 (b), after determining out the minimum cut $\{(v_s, v_t), (v_2, v_t)\}$ in $G'(V',E')$. We have to further construct the minimum cut for $G(V,E)$. Clearly, it is observed that $(v_s, v_t)$ is corresponding to $\{(v_1, v_3), (v_1, v_9), (v_8, v_7)\}$, and $(v_2, v_t)$ corresponds to $(v_2, v_9)$. Therefore, we have $E_{\textrm{opt}}(S, T) = \{(v_1, v_3), (v_1, v_9), (v_2, v_9), (v_8, v_7)\}$.

\textbf{Remark.} For a given directed graph, one can determine the maximum flow from the source vertex to the sink vertex by any efficient maximum flow algorithm. After calling a maximum flow algorithm, the flow information passing through each vertex can be identified. Notice that, the flow passing through an edge will be not larger than its capacity. To determine the cut, one can find the set of vertices that are reachable from the source vertex in the corresponding \emph{residual} graph \cite{algo:2009}. All edges involving a reachable vertex and a non-reachable vertex constitute the minimum cut. This can be completed by applying depth-first search (DFS) technique with a time complexity of $O(|V|+|E|)$.

\section{Steganographic Strategy Between Neighbors}
It seems to be interesting and reasonable that the steganographic communication in a steganographer network is only available between adjacent vertices (or say between neighbors). This scenario looks closer to reality comparing with the case that an encoder in $G$ should participate in the path planning for steganographic communication. The reason is that an edge between any two vertices does not only represent the steganographic channel between them, but also, to a certain extent, shows the social relationship between them. Thus, it may be not desirable sometimes for a vertex to communicate a message along a predetermined network path with another vertex that is not a neighbor. In this section, we will analyze steganography between neighbors within the steganographer network.
\begin{figure}[!t]
\centering
\includegraphics[width=4.5in]{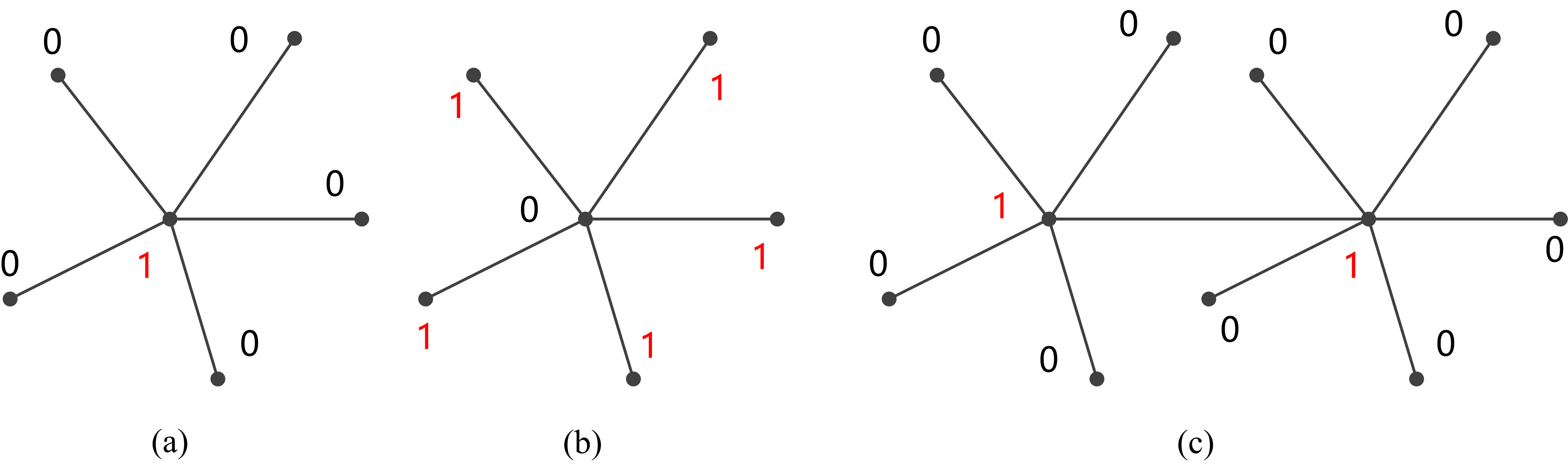}
\caption{Three examples of dividing all vertices into two sets $S$ and $T$: a vertex marked as ``1'' means it belongs to $S$ and ``0'' for $T$.}
\end{figure}

Mathematically, for a given $G(V, E)$, all vertices will serve as either a data encoder or a data decoder. One may assume that, $G(V, E)$ is a subnetwork determined from a complex network in which the vertices may serve as more complex roles. Therefore, it can be said that, we actually deal with a simplified model in this paper even though the complex cases are not explicit for us at present. We are to choose a set of vertices from $V$ to
constitute $S$ mentioned previously and the rest will constitute $T$, i.e., $T=V\setminus S$. We define the neighbor-set of $v\in V$ as $N(v) = \{u| (u,v)\in E\}$. It is required that, for any $v\in T$, there should exist at least one $u\in S$ such that $u\in N(v)$. It ensures that, any vertex in $V$ can either hold a message by itself (in $S$) or receive a message directly from a neighbor belonging to $S$. Accordingly, all vertices can share a message.

The determination of $S$ corresponds to a binary network game \cite{network:games}. Clearly, each individual vertex $v\in V$ must choose an action $g(v)\in X=\{0, 1\}$, where action $g(v)=0$ indicates it servers as a data decoder, and $g(v)=1$ for a data encoder. Therefore, it is required that
\begin{equation}
\forall v\in V, \sum_{u\in N(v)}g(u)\geq 1 - g(v).
\end{equation}

Fig. 5 shows three examples of dividing $V$ into two sets $S$ and $T$. It is seen that, there exist lots of legal solutions. An intuitive requirement is to minimize the number of data encoders, i.e., $\sum_{v\in V}g(v)$, subject to Eq. (9). Simply minimizing $\sum_{v\in V}g(v)$, however, indicates that all vertices essentially have no difference and have the same importance (or say that they have the same risk/cost). From a generalized viewpoint, we can assign a positive weight $w(v)$ to each $v\in V$ to evaluate the cost or risk of marking $v$ as a member of $S$. By assuming an \emph{additive} $G(V,E)$, under the constraint of Eq. (9), our task is to minimize
\begin{equation}
\sum_{v\in V}g(v)\cdot w(v),
\end{equation}
which is a typical 0-1 integer linear programming (ILP) problem.

In graph theory, the family of vertex/edge covering problems involves a broad range of NP-hard optimization problems, among which the minimum-weight dominating set (MWDS) problem has played a prominent role in various real-world domains such as social networks, wireless ad-hoc networks, communication networks, and industrial applications \cite{wang:mwds}. For $G(V, E)$, a dominating set $D$ is a subset of $V$ such that each $v\in V\setminus D$ is adjacent to at least one member of $D$. The MWDS problem aims to find a dominating set $D_\text{min}$ that minimizes the total positive weights assigned to the vertices in the dominating set, namely
\begin{equation}
D_\textrm{min} = \underset{D\subset V}{\textrm{arg min}}~~~\sum_{v\in D}w(v),
\end{equation}
subject to
\begin{equation}
\forall u\in V\setminus D, N(u)\cap D \neq \emptyset.
\end{equation}

Obviously, the ILP problem shown in Eqs. (9, 10) is equivalent to the MWDS problem. Since the MWDS problem is known as a NP-hard problem, generally we have to use approximation algorithms to find the near-optimal solution unless all NP problems can be effectively solved \cite{feige:setcover} or $G(V,E)$ has a small size (or has some special topological structure). For steganography between neighbors in $G$, a core research is therefore to design effective approximation algorithms for the MWDS problem.

\section{Conclusion and Discussion}
In this paper, we introduce two simplified optimization problems in a steganographer network theoretically. Both problems are proven to be equivalent to the minimum cut problem (or say the maximum flow problem since they are dual to each other) and the minimum-weight dominating set problem, respectively. The optimal/near-optimal solutions to the corresponding problems can be found by exploiting graph-modification techniques (e.g., vertex contraction) as well as related deterministic/approximation algorithms used in graph theory.

For the passive attack, to identify the suspicious vertices, one may use steganalysis algorithms or anomaly detection algorithms designed for complex networks. On the one hand, these approaches provide the attacker access to finding both suspicious vertices and edges (corresponding to suspicious channels). On the other hand, the suspicious edges will help the attacker to evaluate the cost of removing an edge from the steganographer network. Notice that, the cost of removing an edge may take into account the local topological characteristics since a steganographer network may contain community features. For example, the edges connecting two communities may be assigned with a high cost. The suited definition of cost of removing an edge will be the future work.

Another reasonable explanation for steganography between neighbors is that, a vertex may have not complete information on the steganographer network. It may lead each vertex to send a message via steganography to its neighbors or receive a message via steganography directly from its neighbors. We propose to find the MWDS in the steganographer network, which, however, has implied that the vertices know the complete information about the network or there should exist an ideal ``\emph{super-vertex}'' that can access the whole network structure and can communicate with each vertex in the network. From the viewpoint of game theory, in case that each vertex has incomplete information about the whole network, the determination of the suited dominating set is more difficult since a vertex may only use its local information and may be affected by its neighbors' actions. Moreover, the determination of weights assigned to vertices in practice is not explicit to us either. We will focus on this problem in future.

\section*{Acknowledgement}
This work was supported by the National Natural Science Foundation of China under Grant Nos. 61502496, U1536120, and U1636201, and the National Key Research and Development Program of China under Grant No. 2016YFB1001003.

\end{document}